\documentclass[aps,prd,onecolumn,groupedaddress,showpacs,nofootinbib,amssymb]{revtex4}
\usepackage[dvips]{graphicx}
\usepackage{amssymb}
\usepackage{amsmath}
\usepackage{graphicx,color}
\usepackage{amsfonts}
\usepackage{bm}
\usepackage{cancel}
\usepackage{comment}

\newcommand\be{\begin{equation}}
\newcommand\ee{\end{equation}}

\allowdisplaybreaks[4]

\begin{document}

\tolerance=5000

\title{Uniqueness of the Inflationary Higgs Scalar for Neutron Stars and Failure of non-inflationary Approximations}
\author{V.K.~Oikonomou,$^{1,2}$\,\thanks{v.k.oikonomou1979@gmail.com}}
 \affiliation{$^{1)}$ Department of Physics, Aristotle University of
Thessaloniki, Thessaloniki 54124,
Greece\\
$^{2)}$ Laboratory for Theoretical Cosmology, Tomsk State
University of Control Systems and Radioelectronics, 634050 Tomsk,
Russia (TUSUR)}

\tolerance=5000

\begin{abstract}
Neutron stars are perfect candidates to investigate the effects of
a modified gravity theory, since the curvature effects are
significant and more importantly, potentially testable. In most
cases studied in the literature in the context of massive
scalar-tensor theories, inflationary models were examined. The
most important of scalar-tensor models is the Higgs model, which,
depending on the values of the scalar field, can be approximated
by different scalar potentials, one of which is the inflationary.
Since it is not certain how large the values of the scalar field
will be at the near vicinity and inside a neutron star, in this
work we will answer the question, which potential form of the
Higgs model is more appropriate in order for it to describe
consistently a static neutron star. As we will show numerically,
the non-inflationary Higgs potential, which is valid for certain
values of the scalar field in the Jordan frame, leads to extremely
large maximum neutron star masses, however the model is not
self-consistent, because the scalar field approximation used for
the derivation of the potential, is violated both at the center
and at the surface of the star. These results shows the uniqueness
of the inflationary Higgs potential, since it is the only
approximation for the Higgs model, that provides self-consistent
results.
\end{abstract}

\pacs{04.50.Kd, 95.36.+x, 98.80.-k, 98.80.Cq,11.25.-w}

\maketitle

\section*{Introduction}

The next two decades will possibly bring sensational observational
results to the cosmology, theoretical physics and theoretical
astrophysics community. All of these observations are related to
gravitational wave detections, either stochastic inflationary
gravitational waves, like the LISA
\cite{Baker:2019nia,Smith:2019wny} and DECIGO
\cite{Seto:2001qf,Kawamura:2020pcg}, or ordinary astrophysical
originating gravitational waves. With regard to astrophysical
sources of gravitational waves, neutron stars are in the epicenter
of current theoretical and experimental research. This is because
neutron stars (NSs)
\cite{Haensel:2007yy,Friedman:2013xza,Baym:2017whm,Lattimer:2004pg,Olmo:2019flu}
are superstars among stars, a wide range of physics research areas
must be used to describe these accurately, like nuclear and high
energy physics
\cite{Lattimer:2012nd,Steiner:2011ft,Horowitz:2005zb,Watanabe:2000rj,Shen:1998gq,Xu:2009vi,Hebeler:2013nza,Mendoza-Temis:2014mja,Ho:2014pta,Kanakis-Pegios:2020kzp,Buschmann:2019pfp,Safdi:2018oeu,Hook:2018iia,Edwards:2020afl,Nurmi:2021xds},
modified gravity can also describe NSs
\cite{Astashenok:2020qds,Capozziello:2015yza,Astashenok:2014nua,Astashenok:2014pua,Astashenok:2013vza,Arapoglu:2010rz,Astashenok:2020cqq,Lobato:2021ehf,Oikonomou:2021iid,Odintsov:2021nqa,Odintsov:2021qbq,Astashenok:2021peo,Astashenok:2021xpm},
and theoretical astrophysics
\cite{Sedrakian:2015krq,Khadkikar:2021yrj,Sedrakian:2006zza,Sedrakian:2018kdm,Bauswein:2020kor,Vretinaris:2019spn,Bauswein:2020aag,Bauswein:2017vtn,Most:2018hfd,Rezzolla:2017aly,Nathanail:2021tay,Koppel:2019pys}.
Four decades passed since the first observation of a NS, and to
date serious questions remain regarding the inner structure and
physics of NSs. The equation of state (EoS) of nuclear matter is
still a mystery in addition to the fundamental question whether
general relativity (GR) or modified gravity
\cite{Nojiri:2017ncd,Capozziello:2011et,Capozziello:2010zz,Nojiri:2006ri,
Nojiri:2010wj,delaCruzDombriz:2012xy,Olmo:2011uz,dimo} controls
the physics of the star. A particularly appealing form of modified
gravity is scalar-tensor gravity, and many works on NSs in the
context of scalar-tensor gravity already exist in the literature
\cite{Pani:2014jra,Staykov:2014mwa,Horbatsch:2015bua,Silva:2014fca,Doneva:2013qva,Xu:2020vbs,Salgado:1998sg,Shibata:2013pra,Arapoglu:2019mun,Ramazanoglu:2016kul,AltahaMotahar:2019ekm,Chew:2019lsa,Blazquez-Salcedo:2020ibb,Motahar:2017blm}.
Also scalar-tensor gravity is popular in cosmological contexts too
\cite{Bezrukov:2014bra,GarciaBellido:2011de,Bezrukov:2010jz,Bezrukov:2007ep,Mishra:2018dtg,Steinwachs:2013tr,Rubio:2018ogq,Kaiser:1994vs,Gundhi:2018wyz,CervantesCota:1995tz,Kamada:2012se,Schlogel:2014jea,Fuzfa:2013yba},
where viable inflationary models can be realized. The model with
the highest importance in scalar-tensor gravity is the Higgs
model, since the Higgs boson is the first fundamental (elementary)
scalar elementary particle that has ever been observed
\cite{Aad:2012tfa}. The Higgs inflationary potential is capable of
producing a viable inflationary era \cite{Mishra:2018dtg} and this
occurs for a specific range of values of the scalar field and the
non-minimal coupling constant to the Ricci scalar, usually denoted
as $\xi$. In a previous work we studied NSs in the context of
scalar-tensor theories, using the inflationary Higgs potential
\cite{Odintsov:2021nqa}. In this work we extend our work to
account for different limiting values of the scalar field and the
combined non-minimal coupling of the form $\sim \xi \phi^2$. We
shall be interested in values $\xi\phi^2\ll 1$ in Geometrized
units. In this approximation, we shall derive the Einstein frame
potential and the relevant conformal transformation function
$A(\phi)$. Accordingly, we shall derive the corresponding
Tolman-Oppenheimer-Volkoff (TOV) equations in the Einstein frame,
for static NSs, and we shall solve these numerically, assuming
piecewise polytropic EoSs \cite{Read:2008iy,Read:2009yp}. We shall
find the $M-R$ relations for static NSs. Our results indicate an
important fact, that the only correct description of the Higgs
potential for static NSs is the one we developed in Ref.
\cite{Odintsov:2021nqa}. The results of the current article
indicate that the maximum masses of NSs exceed the 3$\,M_{\odot}$
limit, but the approximation $ \xi \phi^2\ll 1$ fails to hold true
at the center and at the surface of the NSs. This result indicates
how unique is the inflationary Higgs potential, for providing a
self-consistent neutron star phenomenology.

\section{non-Inflationary Higgs Scalar-tensor Gravity in the Einstein Frame and Static NSs Phenomenology}

We are interested in extracting the Einstein frame counterpart
theory of the Jordan frame Higgs theory, and we shall do so by
using a conformal transformation, see
\cite{Pani:2014jra,Kaiser:1994vs,valerio,Faraoni:2013igs,Buck:2010sv,Faraoni:1998qx}.
for details on conformal transformations. The Jordan frame action
of the Higgs model as it appears in cosmological contexts
\cite{Mishra:2018dtg}, in Geometrized units ($G=1$) is the
following,
\begin{equation}\label{ta}
\mathcal{S}=\int
d^4x\frac{\sqrt{-g}}{16\pi}\Big{[}f(\phi)\mathcal{R}-\frac{1}{2}g^{\mu
\nu}\partial_{\mu}\phi\partial_{\nu}\phi-U(\phi)\Big{]}+S_m(\psi_m,g_{\mu
\nu})\, ,
\end{equation}
where  $f(\phi)$ is the non-minimal coupling function and
$U(\phi)$ is the potential, defined as follows,
\begin{equation}\label{fofphi}
f(\phi)=1+\xi \phi^2,\,\,\, U(\phi)=\lambda \phi^4\, ,
\end{equation}
where $\phi$ denotes the Jordan frame scalar field. Also $g^{\mu
\nu}$, $S_m(\psi_m,g_{\mu \nu})$, $g$ and $\mathcal{R}$ denote the
metric tensor, the action for the matter fluids, the determinant
of the metric tensor and the Ricci scalar in the Jordan frame.

Performing the conformal transformation $\tilde{g}_{\mu
\nu}=A^{-2}g_{\mu \nu}$, where the function $A(\phi)$ is defined
as,
\begin{equation}\label{alphaconformal}
A(\phi)=f^{-1/2}(\phi)\, ,
\end{equation}
and the Einstein frame action in terms of the canonical scalar
field $\varphi$ reads,
\begin{equation}\label{ta5higgs}
\mathcal{S}=\int
d^4x\sqrt{-\tilde{g}}\Big{(}\frac{\tilde{\mathcal{R}}}{16\pi}-\frac{1}{2}
\tilde{g}_{\mu \nu}\partial^{\mu}\varphi
\partial^{\nu}\varphi-\frac{V(\varphi)}{16\pi}\Big{)}+S_m(\psi_m,A^2(\varphi)g_{\mu
\nu})\, ,
\end{equation}
where the ``tilde'' denotes quantities evaluated in the Einstein
frame. Specifically, $\tilde{g}^{\mu \nu}$,
$S_m(\psi_m,A^2(\varphi)g_{\mu \nu})$, $\tilde{g}$ and
$\tilde{\mathcal{R}}$ denote the metric tensor, the action for the
matter fluids, the determinant of the metric tensor and the Ricci
scalar in the Einstein frame.

Recall that $A(\phi)$ enters in the conformal transformation
$\tilde{g}_{\mu \nu}=A^{-2}g_{\mu \nu}$ and by using Eqs.
(\ref{fofphi}) and (\ref{alphaconformal}) we have,
\begin{equation}\label{adef}
 A(\phi)=\left(1+\xi \phi^2 \right)^{-1/2}\, .
\end{equation}
Also the Einstein frame potential is,
\begin{equation}\label{ta3higgs}
V(\phi)=\frac{U(\phi)}{f^2(\phi)}\, ,
\end{equation}
and when expressed in terms of $\phi$ this is written as,
\begin{equation}\label{potinitial}
V(\phi)=\frac{\lambda\phi^4}{\left(1+\xi \phi^2 \right)^2}\, ,
\end{equation}
Using relation between the Einstein frame canonical scalar field
$\varphi$ and the Jordan frame scalar field $\phi$,
\begin{equation}\label{ta4higgs}
\frac{d \varphi }{d \phi}=\frac{1}{\sqrt{4\pi}}
\sqrt{\Big{(}\frac{3}{4}\frac{1}{f^2}\Big{(}\frac{d
f}{d\phi}\Big{)}^2+\frac{1}{4f}\Big{)}}\, ,
\end{equation}
and combined with Eq. (\ref{fofphi}), we get,
\begin{equation}\label{finalrelationdiffphivaphi}
\frac{d \varphi }{d
\phi}=\frac{1}{\sqrt{16\pi}}\frac{\sqrt{1+\xi\phi^2+12\xi^2\phi^2}}{1+\xi\phi^2}\,
.
\end{equation}
For the Higgs inflationary potential, the field values
approximations are the following
\begin{equation}\label{approxmain1alt}
\xi^2\phi^2\gg 1,\,\,\,\xi^2\phi^2\gg \xi \phi^2\, ,
\end{equation}
however, we shall use another approximation relevant in Higgs
potential physics, namely \cite{Mishra:2018dtg},
\begin{equation}\label{approxmain1alt1}
\frac{1}{\sqrt{12}\xi}\ll \phi \ll \frac{1}{\sqrt{\xi}}\, ,
\end{equation}
which is equivalent to the following two approximations,
\begin{equation}\label{approxmain1}
\xi \phi^2\ll 1\, ,
\end{equation}
\begin{equation}\label{approxmain1altaux}
12 \xi^2\phi^2\gg 1\, .
\end{equation}
In view of the approximations (\ref{approxmain1}) and
(\ref{approxmain1altaux}) we get approximately at leading order,
\begin{equation}\label{final1}
\frac{d \varphi }{d \phi}\simeq \frac{\sqrt{12}}{\sqrt{16\pi}}\xi
\phi\, .
\end{equation}
Thus integrating the above we get the final relation between
$\varphi$ and $\phi$,
\begin{equation}\label{finalvarphiphirelation}
\varphi=\frac{\sqrt{12}}{2\sqrt{16\pi}}\xi \phi^2\, .
\end{equation}
At leading order the function $A(\varphi)$ reads,
\begin{equation}\label{alphacapleading}
A(\varphi)=1-\frac{2\sqrt{12}}{\sqrt{16\pi}}\varphi\, ,
\end{equation}
thus at leading order $\alpha (\varphi)=\frac{\rm{d}\ln
A}{\rm{d}\varphi}=-2\sqrt{\frac{16 \pi}{12}}\left(1+ 2
\sqrt{\frac{16 \pi}{12}}\varphi \right)$. Moreover, the potential
as function of $\varphi$ is,
\begin{equation}\label{potentialfinalofvarphi}
V(\varphi)\simeq \lambda \left(\frac{2\sqrt{16
\pi}}{\sqrt{12}}\right)^2\frac{\varphi^2}{\xi^2}\, .
\end{equation}
\begin{figure}[h!]
\centering
\includegraphics[width=20pc]{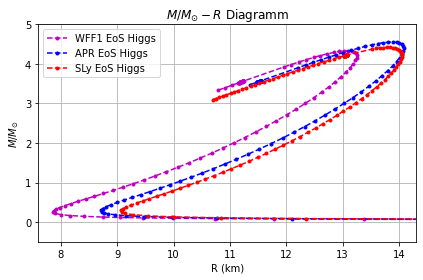}
\caption{$M-R$ graphs for the alternative non-inflationary Higgs
model for the WFF1 EoS (purple curve), the APR EoS (blue curve),
and the SLy EoS (red curve). The $y$-axis is expressed in
$M/M_{\odot}$ units, with $M$ denoting the Jordan frame ADM mass,
and the $x$-axis is the circumferential radius.} \label{plot1}
\end{figure}
For phenomenological reasoning \cite{Mishra:2018dtg}, we shall
choose $\xi \sim 11.455\times 10^4$ with $\lambda=0.1$. With
regard to the EoS, we shall use a piecewise polytropic EoS, the
details of which can be found in \cite{Odintsov:2021nqa}.


For $\xi \sim 11.455\times 10^4$, the requirement
(\ref{approxmain1altaux}) can in principle be satisfied, but the
constraint of Eq. (\ref{approxmain1}) is not necessarily
satisfied. As we will show, this is the case for the
non-inflationary Higgs potential, and we shall verify this
numerically. For the study we shall consider static NSs, which are
described by a spherically symmetric static spacetime of the form,
\begin{equation}\label{tov1}
ds^2=-e^{\nu(r)}dt^2+\frac{dr^2}{1-\frac{2
m}{r}}+r^2(d\theta^2+\sin^2\theta d\phi^2)\, ,
\end{equation}
where $m(r)$ denotes the gravitational mass of the stellar object
confined inside a radius $r$.
\begin{figure}[h!]
\centering
\includegraphics[width=20pc]{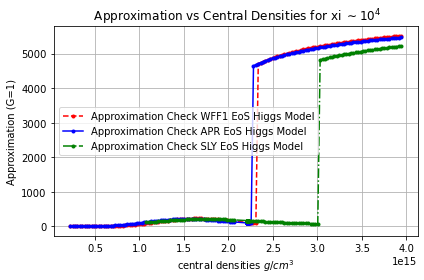}
\caption{The quantity $\xi\phi^2$ ($y$-axis) in Geometrized units,
versus the central densities in CGS units, for $\xi \sim
11.455\times 10^4$, for the WFF1 (red curve), APR (blue curve) and
Sly (green curve) EoSs. As it can be seen the constraint
(\ref{approxmain1}) is not satisfied. } \label{plot2}
\end{figure}
For Geometrized units  ($c=G=1$), the TOV equations for the
spherically symmetric spacetime are,
\begin{equation}\label{tov2}
\frac{d m}{dr}=4\pi r^2 A^4(\varphi)\epsilon+2\pi
r(r-2m)\omega^2+4\pi r^2V(\varphi)\, ,
\end{equation}
\begin{equation}\label{tov3}
\frac{d\nu}{dr}=4\pi r\omega^2+\frac{2}{r(r-2m)}\Big{[}4\pi
A^4(\varphi)r^3P-4\pi V(\varphi) r^3\Big{]}+\frac{2m}{r(r-2m)}\, ,
\end{equation}
\begin{equation}\label{tov4}
\frac{d\omega}{dr}=\frac{r
A^4(\varphi)}{r-2m}\Big{(}\alpha(\varphi)(\epsilon-3P)+4\pi
r\omega(\epsilon-P)\Big{)}-\frac{2\omega
(r-m)}{r(r-2m)}+\frac{8\pi \omega r^2 V(\varphi)+r\frac{d
V(\varphi)}{d \varphi}}{r-2 m}\, ,
\end{equation}
\begin{equation}\label{tov5}
\frac{dP}{dr}=-(\epsilon+P)\Big{[}\alpha (\varphi)\omega+2\pi r
\omega^2+\frac{m-4\pi r^3(-A^4P+V)}{r(r-2m)}\Big{]}\, ,
\end{equation}
\begin{equation}\label{tov5a}
\frac{d\varphi }{dr}=\omega\, ,
\end{equation}
The TOV equations must be solved numerically subject to the
following initial conditions,
\begin{equation}\label{initialconditions}
P(0)=P_c,\,\,\,m(0)=0,\,\,\,\nu(0)\,
,=-\nu_c,\,\,\,\varphi(0)=\varphi_c,\,\,\,\omega (0)=0\, ,
\end{equation}
where $P_c$, $\nu_c$, $\phi_c$ are the pressure of the NS, the
value of the function $\nu(r)$ and the value of the scalar field
at the center of the NS. The values of $nu_c$ and $\varphi_c$ at
the center of the star, shall be obtained using a double shooting
method, in order for the optimal values of them to be obtained.
The requirement for obtaining the optimal values is the scalar
field values to vanish at numerical infinity, which proves to be
the same numerically as in the inflationary Higgs potential,
namely $r\sim 67.94378528694695$ km in the Einstein frame, see
\cite{Odintsov:2021nqa}. Also, for the derivation of the $M-R$
gravity we need to consider the ADM Jordan frame mass and the
Jordan frame radius. Denoting with $r_E$ the Einstein frame radius
at large distances, and $\frac{d\varphi }{dr}=\frac{d\varphi
}{dr}\Big{|}_{r=r_E}$, the Jordan frame mass $M_J\equiv M$ is
related to the Einstein frame mass as follows,
\begin{equation}\label{jordanframeADMmassfinal}
M_J=A(\varphi(r_E))\left(M_E-\frac{r_E^{2}}{2 }\alpha
(\varphi(r_E))\frac{d\varphi
}{dr}\left(2+\alpha(\varphi(r_E))r_E\frac{d \varphi}{dr}
\right)\left(1-\frac{2 M_E}{r_E} \right) \right)\, ,
\end{equation}
with $\frac{d\varphi }{dr}=\frac{d\varphi }{dr}\Big{|}_{r=r_E}$
and $r_J=A r_E$, and $r_J$ is the Jordan frame radius. The
Einstein frame radius $R_s$ of the star can be obtained by the
numerical code by using the condition $P(R_s)=0$, so it is
basically determined by the condition that the pressure of the
star vanishes at the surface of the star. Accordingly, by finding
$R_s$ we can obtain the Jordan frame radius $R$ using the relation
$R=A(\varphi(R_s))\, R_s$, where $\varphi(R_s)$ is the value of
the scalar field at the surface of the star. Finally and important
note is to verify numerically the validity of the approximation
(\ref{approxmain1}) in the Jordan frame. For the numerical
analysis, we shall use a freely available PYTHON code pyTOV-STT
\cite{niksterg}, and we shall derive the solutions for both the
interior and the exterior of the NS, using the ``LSODA'' numerical
method. The EoSs we shall use are the WFF1 \cite{Wiringa:1988tp},
the SLy \cite{Douchin:2001sv}, and the APR EoS
\cite{Akmal:1998cf}. Let us proceed to the results of our
analysis, and we start off with the  $M-R$ graphs for all the EoSs
which we present in Fig. \ref{plot1}. The purple curve corresponds
to the WFF1 EoS, while the red and blue to the SLy and APR EoSs
respectively. From the graphs it is apparent that for the
non-inflationary Higgs model, the maximum masses are comparably
higher with regard to the GR ones. Also in Table \ref{table1} we
present all the maximum masses for all the EoSs corresponding to
the alternative Higgs model. As it can be seen in Table
\ref{table1}, the maximum masses for the alternative Higgs model
are quite elevated compared to the GR ones. Also the GW170817
constraint which indicates that the radius corresponding to the
maximum NS mass must be larger than  is satisfied
$R=9.6^{+0.14}_{-0.03}$km.
\begin{table}[h!]
  \begin{center}
    \caption{\emph{\textbf{Maximum Masses and the of Static NS for the non-Inflationary Higgs Model and for GR}}}
    \label{table1}
    \begin{tabular}{|r|r|r|r|}
     \hline
      \textbf{Model}   & \textbf{APR EoS} & \textbf{SLy EoS} & \textbf{WFF1 EoS}
      \\  \hline
      \textbf{GR} & $M_{max}= 2.18739372\, M_{\odot}$ & $M_{max}= 2.04785291\, M_{\odot}$ & $M_{max}= 2.12603999\, M_{\odot}$
      \\  \hline
      \textbf{Alternative Higgs $\xi\sim 10^4$} & $M_{max}= 4.55374471\,M_{\odot}$ & $M_{max}= 4.41766131\,M_{\odot}$
      &$M_{max}= 4.33460622\, M_{\odot}$ \\  \hline
    \end{tabular}
  \end{center}
\end{table}
The results are deemed quite interesting, however the
non-inflationary Higgs model has inherent issues with the
approximation (\ref{approxmain1}) as we proved numerically.
Particularly it is not satisfied neither at the center nor at the
surface of the star. This feature can be clearly seen in Fig.
\ref{plot2}, where we present the values of $\xi\phi^2$ in the
Jordan frame for all the EoS for the surface scalar field values.
The same applies for the values of the scalar field in the center
of the star. Therefore, to our original question whether
inflationary scalar potentials or other approximations must be
used for static NSs phenomenology, the answer seems to be that
only inflationary potentials provide consistent results.

\section*{Concluding Remarks}

In the field of cosmology there exist several massive scalar field
theories which can potentially play an important role for
describing NSs phenomenology. From these theories, the most
important is the Higgs inflationary theory in its various forms.
Specifically, depending on the scalar field values, the Higgs
potential can take various forms, each of which may describe a
different era in the cosmological theory. Thus the question is
which approximate Higgs potential can describe in a viable and
consistent static NS phenomenology. In this paper we addressed
this question for the most fundamental of all the scalar field
cosmologies, the Higgs inflationary theory. We considered the
theory in the Jordan frame and upon conformally transforming it,
we derived the Einstein frame theory. Accordingly, assuming a
specific range for for the scalar field values, we derived the
appropriate quantities which are relevant for studying static NSs
in the Einstein frame. For a static spherically symmetric
spacetime we derived the TOV equations and we numerically solved
them using a double shooting method for optimizing the results.
The numerical analysis yielded the Einstein frame masses and radii
of the static NS, and also the Einstein frame values of the scalar
field, from which we found the corresponding Jordan frame
quantities. We constructed the $M-R$ graphs and we investigated
the validity of the approximations holding true for the
non-inflationary Higgs model. As we showed, the maximum masses for
the alternative Higgs model are quite elevated, compared with the
GR case, however for all the EoSs studies, the approximation we
assumed for deriving the theory break down. Thus although the
theory provides interesting result, the inherent structure of it
is not correct and consistent. This indicates strongly the
suitability of inflationary potentials for studying NSs
phenomenology, regardless how well motivated other forms of
potentials might be. Moreover, it seems that the approximations
for the scalar field values used for deriving the inflationary
potentials, are well respected on the surface, center and at
numerical infinity of the NS. Hence in conclusion the Higgs
potential that is used for inflationary phenomenology is the only
suitable for describing consistently NSs.

We need to note with regard to the EoSs we used, that we used the
PAR, and more importantly the WFF1, known as FPS EoS, and the SLy,
which both are known to provide a unified description of the crust
and core of NSs. However, all these EoSs are to date rather old
(nearly 20 years old), thus it is compelling to incorporate to the
analysis more timely and to date EoSs, like the BSk24
\cite{Pearson:2018tkr,Pearson:2020bxz}.



\section*{Acknowledgments}

I am grateful to the referee 3 of the manuscript who pointed out
the timely BSk24, which I intend to incorporate in my code for the
piecewise polytropic EoS.


\begin{thebibliography}{999}
\bibitem{Baker:2019nia}
Baker J. \textit{et al.} The Laser Interferometer Space Antenna:
Unveiling the Millihertz Gravitational Wave Sky, [arXiv:1907.06482
[astro-ph.IM]].
\bibitem{Smith:2019wny}
Smith T.~L.; Caldwell R, LISA for Cosmologists: Calculating the
Signal-to-Noise Ratio for Stochastic and Deterministic Sources,
Phys. Rev. D 100 \textbf{(2019)}  no.10, 104055
\bibitem{Seto:2001qf}
Seto N.; Kawamura S.; Nakamura  T., Possibility of direct
measurement of the acceleration of the universe using 0.1-Hz band
laser interferometer gravitational wave antenna in space, Phys.
Rev. Lett. 87 \textbf{(2001)} , 221103
[astro-ph]].
\bibitem{Kawamura:2020pcg}
Kawamura S. \textit{et al.}, Current status of space gravitational
wave antenna DECIGO and B-DECIGO, [arXiv:2006.13545 [gr-qc]].
\bibitem{Haensel:2007yy}
Haensel P.; Potekhin A. Y.; Yakovlev D. G., Neutron stars 1:
Equation of state and structure, Astrophys. Space Sci. Libr.
\textbf{(2007)}, 326, pp.1-619
\bibitem{Friedman:2013xza}
Friedman J. L.; Stergioulas N., Rotating Relativistic Stars,
Cambridge University Press, \textbf{2013}
doi:10.1017/CBO9780511977596
\bibitem{Baym:2017whm}
Baym G.; Hatsuda T.; Kojo T.; Powell P. D.; Song Y.; Takatsuka T.,
From hadrons to quarks in neutron stars: a review, Rept. Prog.
Phys. \textbf{(2018)}, 81 no.5, 056902
\bibitem{Lattimer:2004pg}
Lattimer J. M.; Prakash M., The physics of neutron stars, Science
\textbf{(2004)}, 304, 536-542
\bibitem{Olmo:2019flu}
Olmo G. J.; Rubiera-Garcia D.; Wojnar A., Stellar structure models
in modified theories of gravity: Lessons and challenges, Phys.
Rept. \textbf{(2020)}, 876, 1-75
\bibitem{Lattimer:2012nd}
Lattimer J. M., The nuclear equation of state and neutron star
masses, Ann. Rev. Nucl. Part. Sci. \textbf{(2012)} , 62, 485-515
\bibitem{Steiner:2011ft}
Steiner A. W.; Gandolfi S., Connecting Neutron Star Observations
to Three-Body Forces in Neutron Matter and to the Nuclear Symmetry
Energy, Phys. Rev. Lett. \textbf{(2012)}, 108, 081102
\bibitem{Horowitz:2005zb}
Horowitz C. J.; Perez-Garcia M. A.; Berry  D. K.; Piekarewicz J.,
Dynamical response of the nuclear 'pasta' in neutron star crusts,
Phys. Rev. C \textbf{(2005)}, 72, 035801
\bibitem{Watanabe:2000rj}
Watanabe G.; Iida K.; Sato K., Thermodynamic properties of nuclear
'pasta' in neutron star crusts, Nucl. Phys. A \textbf{(2000)},
676, 455-473 [erratum: Nucl. Phys. A \textbf{726} (2003), 357-365]
doi:10.1016/S0375-9474(00)00197-4 [arXiv:astro-ph/0001273
[astro-ph]].
\bibitem{Shen:1998gq}
Shen H.; Toki H.; Oyamatsu K.; Sumiyoshi K., Relativistic equation
of state of nuclear matter for supernova and neutron star, Nucl.
Phys. A \textbf{(1998)}, 637, 435-450
[nucl-th]].
\bibitem{Xu:2009vi}
Xu J.; Chen L. W.; Li B. A.; Ma H. R., Nuclear constraints on
properties of neutron star crusts, Astrophys. J. \textbf{(2009)},
697, 1549-1568
\bibitem{Hebeler:2013nza}
Hebeler K.; Lattimer J. M.; Pethick C. J.; Schwenk A., Equation of
state and neutron star properties constrained by nuclear physics
and observation, Astrophys. J. \textbf{(2013)}, 773, 11
\bibitem{Mendoza-Temis:2014mja}
 Mendoza-Temis J. de Jes\'us; Wu M. R.; Mart\'\i{}nez-Pinedo G.;
Langanke K.; Bauswein A.; Janka H. T., Nuclear robustness of the r
process in neutron-star mergers, Phys. Rev. C \textbf{(2015)}, 92
no.5, 055805
\bibitem{Ho:2014pta}
Ho W. C. G.; Elshamouty K. G.; Heinke C. O.; Potekhin A. Y., Tests
of the nuclear equation of state and superfluid and
superconducting gaps using the Cassiopeia A neutron star, Phys.
Rev. C \textbf{(2015)}, 91 no.1, 015806
\bibitem{Kanakis-Pegios:2020kzp}
Kanakis-Pegios, Koliogiannis P. S.; Moustakidis C. C., Probing the
nuclear equation of state from the existence of a $\sim
2.6~M_{\odot}$ neutron star: the GW190814 puzzle,
\bibitem{Buschmann:2019pfp}
Buschmann M.; Co R. T.; Dessert C.; Safdi B. R., X-ray Search for
Axions from Nearby Isolated Neutron Stars, Phys. Rev. Lett.
\textbf{(2021)}, 126 no.2, 021102
doi:10.1103/PhysRevLett.126.021102 [arXiv:1910.04164 [hep-ph]].
\bibitem{Safdi:2018oeu}
Safdi B. R.; Sun Z.; Chen A. Y., Detecting Axion Dark Matter with
Radio Lines from Neutron Star Populations, Phys. Rev. D
\textbf{(2019)},99 no.12, 123021
\bibitem{Hook:2018iia}
Hook A.; Kahn Y.; Safdi B. R.; Sun Z., Radio Signals from Axion
Dark Matter Conversion in Neutron Star  Magnetospheres, Phys. Rev.
Lett. \textbf{(2018)},121 no.24, 241102
\bibitem{Edwards:2020afl}
Edwards T. D. P.; Kavanagh B. J.; Visinelli L.; Weniger C.,
Transient Radio Signatures from Neutron Star Encounters with QCD
Axion Miniclusters,
\bibitem{Nurmi:2021xds}
Nurmi S.; Schiappacasse E. D.; Yanagida T. T., Radio signatures
from encounters between Neutron Stars and QCD-Axion Minihalos
around Primordial Black Holes,
\bibitem{Astashenok:2020qds}
Astashenok A. V.; Capozziello S.; Odintsov S. D.; Oikonomou V. K.,
Extended Gravity Description for the GW190814 Supermassive Neutron
Star, Phys. Lett. B \textbf{(2020)}, 811, 135910
\bibitem{Capozziello:2015yza}
Capozziello S.; De Laurentis M.; Farinelli R.; Odintsov S. D.,
Mass-radius relation for neutron stars in f(R) gravity, Phys. Rev.
D \textbf{(2016)}, 93 no.2, 023501
\bibitem{Astashenok:2014nua}
Astashenok A. V.; Capozziello S.; Odintsov S. D., Extreme neutron
stars from Extended Theories of Gravity, JCAP \textbf{(2015)},01,
001
\bibitem{Astashenok:2014pua}
Astashenok A. V.; Capozziello S.; Odintsov S. D., Maximal neutron
star mass and the resolution of the hyperon puzzle in modified
gravity, Phys. Rev. D \textbf{(2014)}, 89 no.10, 103509
\bibitem{Astashenok:2013vza}
Astashenok A. V.; Capozziello S.; Odintsov S. D., Further stable
neutron star models from f(R) gravity, JCAP \textbf{(2013)}, 12,
040
\bibitem{Arapoglu:2010rz}
Arapoglu A. S.; Deliduman C.; Eksi K. Y., Constraints on
Perturbative f(R) Gravity via Neutron Stars, JCAP \textbf{(2011)},
07, 020
\bibitem{Astashenok:2020cqq}
  Astashenok A. V.; Odintsov S. D.,
  Rotating Neutron Stars in F(R) Gravity with Axions,
  Mon.\ Not.\ Roy.\ Astron.\ Soc.\  {\bf(2020) }, 498 no.3, 3616
\bibitem{Lobato:2021ehf}
Lobato R.~V.; Carvalho G.~A.; Bertulani C.~A., Neutron stars in
$f(\mathtt{R,L_m})$ gravity with realistic equations of state:
joint-constrains with GW170817, massive pulsars, and the PSR
J0030+0451 mass-radius from ${\it NICER}$ data, Eur. Phys. J. C
\textbf{(2021)}, 81, 1013
[arXiv:2106.01841 [gr-qc]].
\bibitem{Oikonomou:2021iid}
Oikonomou V.~K., Universal inflationary attractors implications on
static neutron stars, Class. Quant. Grav. \textbf{(2021)}, 38
no.17, 175005
[gr-qc]].
\bibitem{Odintsov:2021nqa}
Odintsov S.~D.; Oikonomou V.~K., Neutron Stars in Scalar-tensor
Gravity with Higgs Scalar Potential, [arXiv:2104.01982 [gr-qc]].
\bibitem{Odintsov:2021qbq}
Odintsov S.~D.; Oikonomou V.~K., Neutron stars phenomenology with
scalar\textendash{}tensor inflationary attractors, Phys. Dark
Univ. \textbf{(2021)}, 32, 100805
[arXiv:2103.07725 [gr-qc]].
\bibitem{Astashenok:2021peo}
Astashenok A.~V.; Capozziello S.; Odintsov S.~D.; Oikonomou V.~K.,
Causal limit of neutron star maximum mass in $f(R)$ gravity in
view of GW190814, Phys. Lett. B \textbf{(2021)}, 816, 136222
doi:10.1016/j.physletb.2021.136222 [arXiv:2103.04144 [gr-qc]].
\bibitem{Astashenok:2021xpm}
Astashenok A.~V.; Capozziello S.; Odintsov S.~D.; Oikonomou V.~K.,
Novel stellar astrophysics from extended gravity, EPL
\textbf{(2021)}, 134 no.5, 59001
[arXiv:2106.01234 [gr-qc]].
\bibitem{Sedrakian:2015krq}
Sedrakian A., Axion cooling of neutron stars, Phys. Rev. D
\textbf{(2016)}, 93 no.6, 065044
\bibitem{Khadkikar:2021yrj}
  Khadkikar S.; Raduta A. R., Oertel M.; Sedrakian A.,
  arXiv:2102.00988 [astro-ph.HE].
\bibitem{Sedrakian:2006zza}
Sedrakian D. M.; Hayrapetyan M. V.; Shahabasyan M. K.,
Gravitational radiation of slowly rotating neutron stars,
Astrophysics \textbf{(2006)}, 49, 194-200
\bibitem{Sedrakian:2018kdm}
Sedrakian A., Axion cooling of neutron stars. II. Beyond hadronic
axions, Phys. Rev. D \textbf{(2019)},99 no.4, 043011
\bibitem{Bauswein:2020kor}
Bauswein A.; Guo G.; Lien J. H., Lin Y. H.; Wu M. R., Compact Dark
Objects in Neutron Star Mergers,
\bibitem{Vretinaris:2019spn}
Vretinaris S.; Stergioulas N.; Bauswein A., Empirical relations
for gravitational-wave asteroseismology of binary neutron star
mergers, Phys. Rev. D \textbf{(2020)},101 no.8, 084039
\bibitem{Bauswein:2020aag}
BausweinA.; Blacker S.; Vijayan V.; Stergioulas N.; Chatziioannou
K.; Clark J. A.; Bastian N. U. F.; Blaschke D. B.; Cierniak M.;
Fischer T., Equation of state constraints from the threshold
binary mass for prompt collapse of neutron star mergers, Phys.
Rev. Lett. \textbf{(2020)},125 no.14, 141103
\bibitem{Bauswein:2017vtn}
Bauswein A.; Just O.; Janka H. T.; Stergioulas N., Neutron-star
radius constraints from GW170817 and future detections, Astrophys.
J. Lett. \textbf{(2017)},850 no.2, L34
\bibitem{Most:2018hfd}
Most E. R., Weih L. R., Rezzolla L.; Schaffner-Bielich J., New
constraints on radii and tidal deformabilities of neutron stars
from GW170817, Phys. Rev. Lett. \textbf{(2018)}, 120 no.26, 261103
\bibitem{Rezzolla:2017aly}
Rezzolla L.; Most E. R.; Weih L. R., Using gravitational-wave
observations and quasi-universal relations to constrain the
maximum mass of neutron stars, Astrophys. J. Lett.
\textbf{(2018)},852 no.2, L25
\bibitem{Nathanail:2021tay}
Nathanail A.; Most E. R.; Rezzolla L., GW170817 and GW190814:
tension on the maximum mass, Astrophys. J. Lett.
\textbf{(2021)},908 no.2, L28
\bibitem{Koppel:2019pys}
K\"oppel S.; Bovard L.; RezzollaL., A General-relativistic
Determination of the Threshold Mass to Prompt Collapse in Binary
Neutron Star Mergers, Astrophys. J. Lett. \textbf{(2019)},872
no.1, L16
\bibitem{Nojiri:2017ncd}
Nojiri S.; Odintsov S. D.; Oikonomou V. K.,
Phys.\ Rept.\  \textbf{(2017)}, 692, 1
\bibitem{Capozziello:2011et}
Capozziello S.; De Laurentis M., Extended Theories of Gravity,
Phys.\ Rept.\  \textbf{(2011)}, 509, 167
\bibitem{Capozziello:2010zz}
Faraoni V.; Capozziello S., Beyond Einstein Gravity : A Survey of
Gravitational Theories for Cosmology and Astrophysics, Fundam.\
Theor.\ Phys.\  \textbf{(2010)},170.
\bibitem{Nojiri:2006ri}
Nojiri S.; Odintsov S. D., Introduction to modified gravity and
gravitational alternative for dark energy, eConf C
\textbf{(2006)}, 0602061, 06
 [Int.\ J.\ Geom.\ Meth.\ Mod.\ Phys.\  \textbf{(2007)}, 4, 115]
\bibitem{Nojiri:2010wj}
Nojiri S.; Odintsov S. D., Unified cosmic history in modified
gravity: from F(R) theory to Lorentz non-invariant models, Phys.\
Rept.\  {(2011)}, \textbf{505}, 59
\bibitem{delaCruzDombriz:2012xy}
de la Cruz-Dombriz A.; Saez-Gomez D., Black holes, cosmological
solutions, future singularities, and their thermodynamical
properties in modified gravity theories, Entropy \textbf{2012)},
14, 1717
\bibitem{Olmo:2011uz}
Olmo G. J., Palatini Approach to Modified Gravity: f(R) Theories
and Beyond, Int.\ J.\ Mod.\ Phys.\ D  \textbf{(2011)}, 20, 413
\bibitem{dimo} Konstantinos Dimopoulos, Introduction to Cosmic Inflation and Dark Energy, (2021) CRC Press
\bibitem{Pani:2014jra}
Pani P.; Berti E., Slowly rotating neutron stars in scalar-tensor
theories, Phys. Rev. D \textbf{(2014)},90 no.2, 024025
\bibitem{Staykov:2014mwa}
Staykov K. V.; Doneva D. D.; Yazadjiev S. S.; Kokkotas K. D.,
Slowly rotating neutron and strange stars in $R^2$ gravity, JCAP
\textbf{(2014)}, 10, 006
\bibitem{Horbatsch:2015bua}
Horbatsch M.; Silva H. O.; Gerosa D.; Pani P.; Berti E., Gualtieri
L.; Sperhake U., Tensor-multi-scalar theories: relativistic stars
and 3 + 1 decomposition, Class. Quant. Grav. \textbf{(2015)},32
no.20, 204001
\bibitem{Silva:2014fca}
Silva H.O.; Macedo C. F. B.; Berti E.; Crispino L. C. B., Slowly
rotating anisotropic neutron stars in general relativity and
scalar\textendash{}tensor theory, Class. Quant. Grav.
\textbf{(2015)}, 32, 145008
\bibitem{Doneva:2013qva}
Doneva D. D.; Yazadjiev S. S.; Stergioulas N.; Kokkotas K. D.,
Rapidly rotating neutron stars in scalar-tensor theories of
gravity, Phys. Rev. D \textbf{(2013)}, 88 no.8, 084060
\bibitem{Xu:2020vbs}
Xu R.; Gao Y.; Shao L., Strong-field effects in massive
scalar-tensor gravity for slowly spinning neutron stars and
application to X-ray pulsar pulse profiles, Phys. Rev. D
\textbf{(2020)},102 no.6, 064057
\bibitem{Salgado:1998sg}
Salgado M.; Sudarsky D.; Nucamendi U., On spontaneous
scalarization, Phys. Rev. D \textbf{(1998)}, 58, 124003
\bibitem{Shibata:2013pra}
Shibata M.; Taniguchi K.; Okawa H.; Buonanno A., Coalescence of
binary neutron stars in a scalar-tensor theory of gravity, Phys.
Rev. D \textbf{(2014)}, 89 no.8, 084005
\bibitem{Arapoglu:2019mun}
Sava\c{s} Arapo\u{g}lu A.; Yavuz Ek\c{s}i K.; Emrah Y\"ukselci A.,
Neutron star structure in the presence of nonminimally coupled
scalar fields, Phys. Rev. D \textbf{(2019)}, 99 no.6, 064055
\bibitem{Ramazanoglu:2016kul}
Ramazano\u{g}lu F. M.; Pretorius F., Spontaneous Scalarization
with Massive Fields, Phys. Rev. D \textbf{(2016)},93 no.6, 064005
\bibitem{AltahaMotahar:2019ekm}
Altaha Motahar Z.; Bl\'azquez-Salcedo J. L., Doneva D. D., Kunz
J.; Yazadjiev S. S., Axial quasinormal modes of scalarized neutron
stars with massive self-interacting scalar field, Phys. Rev. D
\textbf{(2019)}, 99 no.10, 104006
\bibitem{Chew:2019lsa}
Chew X. Y.; Dzhunushaliev V.; Folomeev V.; Kleihaus B.; Kunz J.,
Rotating wormhole solutions with a complex phantom scalar field,
Phys. Rev. D \textbf{(2019)}, 100 no.4, 044019
\bibitem{Blazquez-Salcedo:2020ibb}
Bl\'azquez-Salcedo J. L.; Scen Khoo F.; Kunz J., Ultra-long-lived
quasi-normal modes of neutron stars in massive scalar-tensor
gravity, EPL \textbf{(2020)}, 130 no.5, 50002
\bibitem{Motahar:2017blm}
Altaha Motahar Z.; Bl\'azquez-Salcedo J. L.; Kleihaus B.; Kunz J.,
Scalarization of neutron stars with realistic equations of state,
Phys. Rev. D \textbf{(2017)},96 no.6, 064046
\bibitem{Bezrukov:2014bra}
Bezrukov F.; Shaposhnikov M., Higgs inflation at the critical
point, Phys. Lett. B \textbf{(2014)}, 734, 249-254
\bibitem{GarciaBellido:2011de}
Garcia-Bellido J.; Rubio J.; Shaposhnikov M.; Zenhausern D.,
Higgs-Dilaton Cosmology: From the Early to the Late Universe,
Phys. Rev. D \textbf{(2011)}, 84, 123504
\bibitem{Bezrukov:2010jz}
Bezrukov F.; Magnin A.; Shaposhnikov M.; Sibiryakov S., Higgs
inflation: consistency and generalisations, JHEP \textbf{(2011)},
01, 016
\bibitem{Bezrukov:2007ep}
Bezrukov F. L.; Shaposhnikov M., The Standard Model Higgs boson as
the inflaton, Phys. Lett. B \textbf{(2008)}, 659, 703-706
\bibitem{Mishra:2018dtg}
Mishra S. S., Sahni V.; Toporensky A. V., Initial conditions for
Inflation in an FRW Universe, Phys. Rev. D \textbf{(2018)}, 98
no.8, 083538
\bibitem{Steinwachs:2013tr}
Steinwachs C. F.; Kamenshchik A. Y., Non-minimal Higgs Inflation
and Frame Dependence in Cosmology, AIP Conf. Proc.
\textbf{(2013)},1514 no.1, 161-164
\bibitem{Rubio:2018ogq}
Rubio J., Higgs inflation, Front. Astron. Space Sci.
\textbf{(2019)}, 5, 50
\bibitem{Kaiser:1994vs}
Kaiser D. I., Primordial spectral indices from generalized
Einstein theories, Phys. Rev. D \textbf{52} (1995), 4295-4306
\bibitem{Gundhi:2018wyz}
Gundhi A.; Steinwachs C. F., Scalaron-Higgs inflation, Nucl. Phys.
B \textbf{(2020)}, 954, 114989
\bibitem{CervantesCota:1995tz}
Cervantes-Cota J. L.; Dehnen H., Induced gravity inflation in the
standard model of particle physics, Nucl. Phys. B \textbf{(1995)},
442, 391-412
\bibitem{Kamada:2012se}
Kamada K.; Kobayashi T.; Takahashi T.; Yamaguchi M.; Yokoyama J.,
Generalized Higgs inflation, Phys. Rev. D \textbf{(2012)}, 86,
023504 doi:10.1103/PhysRevD.86.023504 [arXiv:1203.4059 [hep-ph]].
\bibitem{Schlogel:2014jea}
Schlogel S.; Rinaldi M.; Staelens F.; Fuzfa A., Particlelike
solutions in modified gravity: the Higgs monopole, Phys. Rev. D
\textbf{(2014)}, 90 no.4, 044056
\bibitem{Fuzfa:2013yba}
F\"uzfa A.; Rinaldi M.; Schl\"ogel S., Particlelike distributions
of the Higgs field nonminimally coupled to gravity, Phys. Rev.
Lett. \textbf{(2013)}, 111 no.12, 121103
\bibitem{Aad:2012tfa}
Aad G.; \textit{et al.} [ATLAS], Observation of a new particle in
the search for the Standard Model Higgs boson with the ATLAS
detector at the LHC, Phys. Lett. B \textbf{(2012)}, 716, 1-29
doi:10.1016/j.physletb.2012.08.020 [arXiv:1207.7214 [hep-ex]].
\bibitem{Read:2008iy}
Read J. S.; Lackey B. D.; Owen B. J. ; Friedman J. L., Constraints
on a phenomenologically parameterized neutron-star equation of
state, Phys. Rev. D \textbf{(2009)}, 79, 124032
\bibitem{Read:2009yp}
Read J. S.; Markakis C.; Shibata M.; Uryu K.; Creighton J. D. E.;
Friedman J. L., Measuring the neutron star equation of state with
gravitational wave observations, Phys. Rev. D \textbf{(2009)}, 79,
124033
\bibitem{valerio} Valerio Faraoni, Cosmology in Scalar-Tensor Gravity,
Springer 2004
\bibitem{Faraoni:2013igs}
Faraoni V., Conformally coupled inflation, Galaxies
\textbf{(2013)}, 1 no.2, 96-106
\bibitem{Buck:2010sv}
Buck M.; Fairbairn M.; Sakellariadou M., Inflation in models with
Conformally Coupled Scalar fields: An application to the
Noncommutative Spectral Action, Phys. Rev. D \textbf{(2010)}, 82,
043509
\bibitem{Faraoni:1998qx}
  Faraoni V.; Gunzig E.; Nardone P.,
  Conformal transformations in classical gravitational theories and in cosmology,
  Fund.\ Cosmic Phys.\  {\bf (1999)}, 20,  121
\bibitem{niksterg} Nikolaos Stergioulas, https://github.com/niksterg
\bibitem{Wiringa:1988tp}
Wiringa R. B.; Fiks V.; Fabrocini A., Equation of state for dense
nucleon matter, Phys. Rev. C \textbf{(1988)}, 38, 1010-1037
\bibitem{Douchin:2001sv}
Douchin F.; Haensel P., A unified equation of state of dense
matter and neutron star structure, Astron. Astrophys.
\textbf{(2001)}, 380, 151
\bibitem{Akmal:1998cf}
Akmal A.; Pandharipande V.~R.; Ravenhall D.~G., The Equation of
state of nucleon matter and neutron star structure, Phys. Rev. C
\textbf{(1998)}, 58, 1804-1828
[arXiv:nucl-th/9804027 [nucl-th]].
\bibitem{Pearson:2018tkr}
Pearson J.~M.; Chamel N.; Potekhin A.~Y.; Fantina A.~F.; Ducoin
C.; Dutta A.~K.;Goriely S.,
Mon. Not. Roy. Astron. Soc. \textbf{(2018)} 481  no.3, 2994-3026
[erratum: Mon. Not. Roy. Astron. Soc. \textbf{(2019)} 486 no.1,
768] doi:10.1093/mnras/sty2413 [arXiv:1903.04981 [astro-ph.HE]].
\bibitem{Pearson:2020bxz}
Pearson J.~M.; Chamel N.; Potekhin A.~Y.~,
Phys. Rev. C \textbf{(2020)} 101 no.1, 015802
doi:10.1103/PhysRevC.101.015802 [arXiv:2001.03876 [astro-ph.HE]].

\end{thebibliography}
\end{document}